\documentclass[12pt]{aastex631}
\usepackage{hhline}
\usepackage{lineno}

\begin{document}

\title{An unusual pair of interstellar HI features and a related white dwarf star inside the HI cavity surrounding the Upper Sco-Cen OB2 Association.}

\author[0000-0002-6160-1040]{Gerrit L. Verschuur}
\affiliation{verschuur@aol.com}

\keywords{ISM: atoms - ISM: clouds, Stars: White dwarfs; Nebulae: Planetary nebulae}

\begin{abstract}
Two mysterious unresolved HI structures at velocities of +12 and -6 km s$^{-1}$ were discovered in high resolution $\lambda$21-cm survey data in the direction of a faint white dwarf star.  Examination of the HI morphology in this area of sky shows that the star and HI features exists in a large cavity in interstellar HI surrounding the Upper Sco-Cen OB2 Association.  The cavity may have been created by an ancient supernova.  It is hypothesized that the pair of HI features and filamentary HI structure found in its immediate vicinity may be the remnants of a planetary nebula some 3 x 10$^{5}$ years old that have cooled to the point that the gas is neutral and emitting the $\lambda$21-cm spectral line.  This remnant has maintained the morphological characteristics of the original planetary nebula because it expanded into a volume of space relatively devoid of interstellar gas that would otherwise have absorbed any traces of the original nebula.  
\end{abstract}

\section{Introduction}

During a study of interstellar neutral hydrogen (HI) structure an apparently unresolved HI cloud at a velocity of +12 km s$^{-1}$ was found at the location of a white dwarf star.  Then a twin HI feature, at -6 km s$^{-1}$,  was discovered at the same location.  These coincidences led to the present work that revealed that  the features are located inside a region of reduced HI emission that encompasses the Upper Branch of the Sco-Cen OB2 Association.   

The physics of the Sco-Cen OB2 Association and its substructure has been the focus of  great deal of research over the past many decades.  It is not our intention to review the literature, but a few references deserve mention in the context of the work to be reported here.  de Geus (1992) in his seminal work ''Interactions of stars and interstellar matter in Scorpio Centaurus'' noted that this is the youngest subgroup of  the much larger Sco-Cen OB2 Association and that it contains a considerable amount of molecular material.  He estimated its total mass to be about 3 x 10$^{3}$ M$_{\circ}$.   Krause et al. (2018) refer to the group of OB stars described by de Geus (1992) as the Upper Scorpius (USco) Association and estimated that it contains a stellar mass of about 2 x 10$^{3}$ M$_{\circ}$.  Olano et al. (1981) detected evidence for a shell of neutral hydrogen around the Upper Sco-Cen association at +8 km/s, see their Fig. 3d, and  de Geus (1992) in his Fig. 2 sketched the HI column density in the same feature, as did  P{\"o}ppel et al. (2010) in their Figs. 3 \& 4, which they labelled as the US-Loop.  

In the present study the highest resolution HI data now available have been used for a closer look at the area in an endeavor to understand the apparent relationship between of the two HI  features and the white dwarf star, which are located in the direction of the HI cavity surrounding the Upper Sco-Cen OB Association.  In what follows the appellation USco will be used refer to this area.

\section{The data}
Data regarding the structure of interstellar neutral hydrogen (HI) were drawn from the \textit{HI4PI} all-sky HI survey (beamwidth 16.\arcmin1, bandwidth 1.28 km s$^{-1}$, sampling interval 0.\arcdeg0833) of Ben Bekhti et al. (2016), which combined the northern hemisphere data from the  Effelsberg-Bonn HI Survey (\textit{EBHIS}; Kerp et al. 2011; Winkel et al. 2016) with the southern sky \textit{GASS} survey data of McClure-Griffiths et al. (2009).  The Leiden-Argentine-Bonn \textit{LAB} survey (beamwidth 36.\arcmin0, bandwidth 1.0 km s$^{-1}$, sampling interval 0.\arcdeg5) from Kalberla et al. (2005) was also drawn upon in the present work.  

The \textit{HI4PI} data were used to produce maps at 2 km s$^{-1}$  intervals, each encompassing only one bandwidth, 1.28 km s$^{-1}$, between -70 and + 70 km s$^{-1}$ and bounded by Galactic latitudes \textit{b} = $\pm$ 85\arcdeg\ and longitude \textit{l} = 260\arcdeg\ through  0\arcdeg\ to 20\arcdeg.  Fig. 1 shows a section of the map at +12 km s$^{-1}$, which offers a context for our study of the HI structure in this area of sky.  A dashed red circle is overlain on the ring of HI that marks the boundary of the HI cavity surrounding the USco Association.  Also marked in Fig, 1 is a red dashed-line rectangle that defines the extent of the Ophiuchus region described by de Geus \& Burton (1991) within which area de Geus et al. (1990) reported results of a CO survey.  There is an overlap between the Ophiuchus and USco boundaries, as is evident in this plot.  

\begin{figure}
\figurenum{1}
\epsscale{1.0}
\plotone{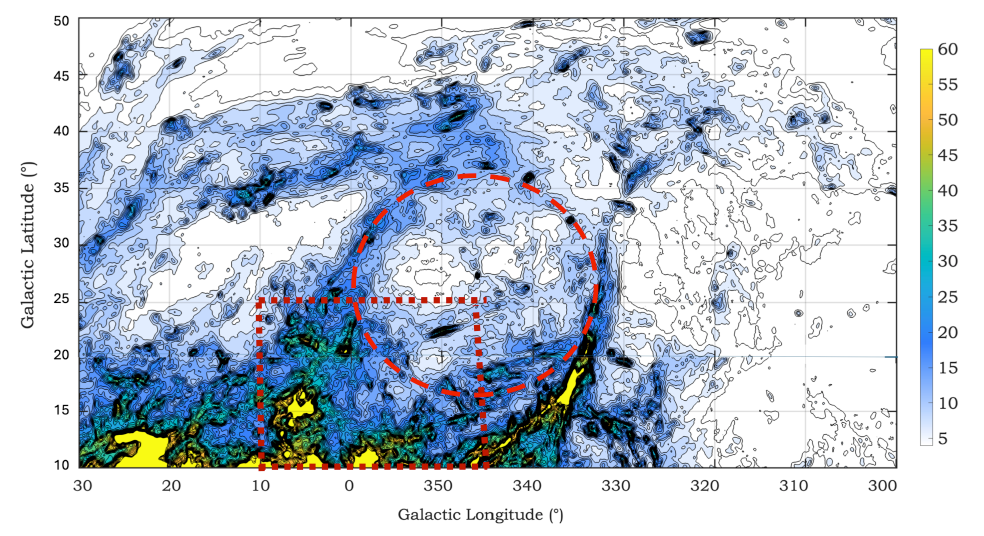}
\caption{The \textit{HI4PI} single channel \textit{l-b} map at +12 km s$^{-1}$ showing the ring of HI surrounding the USco Association in the center of the area outlined by the red dashed circle that extends from about \textit{l} = 333\arcdeg to 0\arcdeg and \textit{b} = 15\arcdeg\ to 35\arcdeg.  The area inside the ring is relatively devoid of HI at this velocity as discussed in the text.   The boundaries  of the Ophiuchus area discussed in the text are indicated  by a red dashed-line rectangle.  The small HI peak associated with a white dwarf star at (\textit{l, b})  = (346.\arcdeg13, +27.\arcdeg38) is located just above the upper right corner of the rectangle.  Legend for the HI brightness is in K km s$^{-1}$. }
\end{figure}

Fig. 1 shows a bright, point-like spot just above the top right-hand corner of the sketched rectangle. That is the mystery HI feature that initially drew our attention.  It is located in an area of very low HI emission at this velocity.  Fig. 2a is a close-up of this structure at +12 km s$^{-1}$ and it appears to be unresolved.  Examination of the full set of \textit{l-b} HI maps led to the discovery of a second apparently unresolved feature, this one at -6 km s$^{-1}$ at essentially the same location.  It is shown in Fig. 2b.  This feature is superimposed on a low-velocity background of order 50 K.km s$^{-1}$.  In both figures the location of the white dwarf star discussed in the next section is indicated.

\begin{figure}
\figurenum{2}
\epsscale{1.0}
\plotone{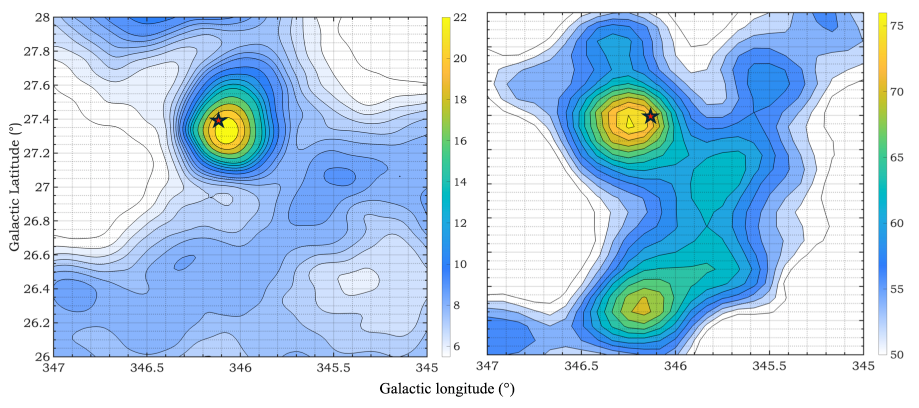}
\caption{a)  A close-up view of the mystery features (Table 1) in a \textit{l-b} map at (\textit{l, b}) = (346.\arcdeg08, 27.\arcdeg33) in a \textit{HI4PI} one channel (1.28 km s$^{-1}$) wide map at a velocity of +12 km s$^{-1}$.  The apparent point source  has the appearance of a beam shape with a small asymmetry to the north.   b) The same area at -6 km s$^{-1}$ that shows an HI peak at (\textit{l, b}) = (346.\arcdeg25, 27.\arcdeg33).  The location of the white dwarf star discussed in the text is indicated by the star symbol.  The nature of other structure in both maps is discussed in \S3.  Legends are in unit of K. km s$^{-1}$.}
\end{figure}

To rule out the possibility that interference caused the  unresolved peaks in the \textit{l-b} maps, Fig. 3a shows the \textit{HI4PI} emission profile at the position of the HI peak, and Fig. 3b is the profile obtained in the same direction using \textit{LAB} survey data.  Both plots reveal a peak around +12 km s$^{-1}$.  Given that the two telescopes involved were located on two continents (Argentina for \textit{LAB}, and Australia for \textit{HI4PI}) this ruled out interference as the cause of the signal.   

\begin{figure}
\figurenum{3}
\epsscale{1.0}
\plotone{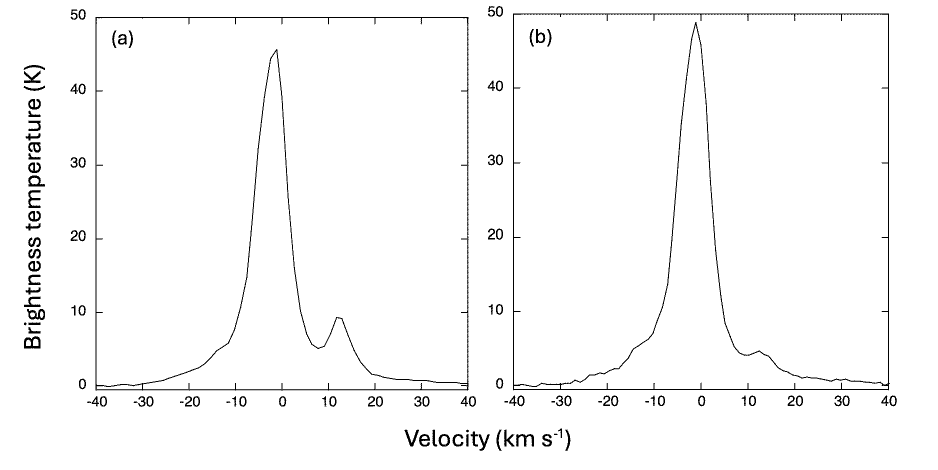}
\caption{Neutral hydrogen emission profiles from two surveys.  a)  \textit{HI4PI} data at the position of the mystery source at (\textit{l, b}) = (346.\arcdeg08, 27.\arcdeg33).   b)  The closest \textit{LAB} profile at (\textit{l, b}) = (346.\arcdeg0, 27.\arcdeg5) shows the same component at +12 km s$^{-1}$.}
\end{figure}

\subsection{Searching SIMBAD}
The SIMBAD data base\footnote{https://simbad.u-strasbg.fr/simbad/sim-fcoo} was queried for objects near the location of the mystery HI features.  At the location of the  HI centers three White Dwarf (WD) candidates were found.  Subsequent examination of the study by Gentile-Fusillo et al. (2021) showed that two of these candidates are in fact white dwarfs.   Table 1 lists their coordinates together with their distances as well as location of the HI peaks and their velocities.  The location of WD \#1 in Table 1 (distance 110 pc) is within 0.\arcdeg1 of the two HI peaks (see Fig. 2) while WD \#2 is located 0.\arcdeg7 in longitude away from the HI peaks.  The apparent association between the HI features and WD \#1 is therefore favored.  
\begin{deluxetable}{cccccc}
\tabletypesize{\scriptsize}	
\tablecolumns{6}	
\tablewidth{0pc}	
\tablecaption{ Properties of Objects of Interest}	
\tablehead{\colhead{Type of object}	&	\colhead{Identifier}	&	\colhead{Galactic longitude}	&	\colhead{Galactic latitude}	&	\colhead{Distance}	&	\colhead{Velocity}	\\ {}	&	{}	&	(\arcdeg)	&	(\arcdeg)	&	(parsecs)	 &	(km s$^{-1}$)	}		
\startdata
HI feature	&	Far side (see text)	&	 346.08	&	27.33	&	-	&	12	\\													
HI feature	&	Near side (see text)	&	 346.25	&	27.33	&	-	&	-6	\\													
White dwarf  \#1	&	GAIA DR3 6253301413213234688	&	 346.13	&	27.38	&	110	&	?	\\													
White dwarf  \#2	&	GAIA DR3 6253281175328656356	&	 345.94	&	27.50	&	334	&	?	\\													
\enddata																								
\end{deluxetable}

\subsection{The angular size of the HI features}
In both Figs. 2a and 2b the mystery peaks appeared at first glance to be unresolved.  In order to determine their true angular extent, every HI profile in the \textit{HI4PI} data base was downloaded  as a function of two cross-sections, one along longitude and the other in latitude across each feature. Knowing the beam width, $\theta$$_{beam}$, the observed angular width, $\theta$$_{obs}$, can be used to calculate the true source width, $\theta$$_{source}$, of the HI inside the beam, assuming it has a Gaussian cross-section.  It can be derived from the formula:
\begin{equation}
\theta_{obs}^2 = \theta_{beam}^2 + \theta_{source}^2.
\end{equation}

Table 2 summarizes the derived values for the two structures. The longitude widths take into account the cosine(latitude) correction and the distance is assumed to be 110 pc, that of the star.  The HI features have angular widths less than the beam width of the \textit{HI4PI} observations.
\begin{deluxetable}{ccc}																																											
\tablecolumns{3}																								
\tablewidth{0pc}																								
\tablecaption{ Sizes of the HI Features}																								
\tablehead{\colhead{Property}	&	\colhead{Longitude X-section}	&	\colhead{Latitude X-section}	}								
\startdata																								
NEAR-SIDE (-6 km s$^{-1}$)	& 	{}	&	{}		\\																	
$\theta$$_{obs}$	&	26.\arcmin8	&	22.\arcmin6	\\
\newline
$\theta$$_{source}$  &    21.\arcmin5      & 15.\arcmin9 \\	
\newline
Linear width  & 0.69 pc   & 0.51 pc\\	
\newline\\
FAR-SIDE (+12 km s$^{-1}$)	& 	{}	&	{}		\\	
$\theta$$_{obs}$	&	18.\arcmin0	&	18.\arcmin4	\\
\newline
$\theta$$_{source}$  &  8.\arcmin0      & 10.\arcmin8 \\	
\newline
Linear width  & 0.26 pc   & 0.35 pc\\	
\enddata																								
\end{deluxetable}																															

\subsection{The physical properties of the two HI features}
Table 3 summarizes the physical parameters describing the structure at +12 km s$^{-1}$ shown in Fig. 2a.  Given the apparent association with the WD, this is interpreted as HI at the far-side of the star.  The HI column density is derived from a Gaussian analysis of the spectrum.  HI emission from the far-side feature is a single component well separated from the bulk of the emission at low velocities, see Fig. 3a.  This makes a Gaussian decomposition unambiguous.  The same is not true for the near-side component at -6 km s$^{-1}$, which is found atop the low velocity HI peak.  Due to the complexity of the low-velocity emission profile, no consistent set of Gaussian parameters could be assigned to this component. Therefore, Table 3 lists only the key physical parameters derived for the far-side (+12 km s$^{-1}$) feature.

\begin{deluxetable}{cc}			
			
\tablecolumns{2}			
\tablewidth{0pc}			
\tablecaption{Properties of the far side HI feature}			
\tablehead{\colhead{Parameter}	&	\colhead{Value}	}
\startdata			
Ave. angular width (FWHM) 	&	0.\arcdeg26	\\
Average radius 	&	0.15	pc\\
Emission line width 	&	4.7 km s$^{-1}$\\
Velocity 	&	12.7	 km s$^{-1}$   \\
HI column density	&	30.2	10$^{18}$ cm$^{-2}$\\
Volume density, but see text	&	32.4 cm$^{-3}$	\\
Mass  &	0.01M$_{\circ}$	\\
Energy &	7  x 10$^{42}$	ergs\\
\enddata			
\end{deluxetable}																			
																						\clearpage										

The apparent volume density is derived by assuming that the observed HI feature is a cloud that is as deep as it is wide.  However, it is likely that the feature is elongated along the line-of-sight by some factor, in which case the derived volume density is correspondingly smaller by that same factor.  The total mass does not change.  Figs. 2a \& b suggests that the two structures seen in Fig. 2 are sufficiently similar to one another that they may have similar physical properties, which makes their combined mass 0.02 M$_{\circ}$.

\subsection{The Environment of the Mystery HI clouds and the White Dwarf}
The mystery pair of HI features and the WD are located in the direction of the cavity in the HI surrounding the USco Association as seen in Fig. 1.  This alignment is inferred from a two-dimensional map but that does not prove that the star and the mystery HI features lie within the cavity in depth.  Also, what caused the cavity in the first place?  If the cavity and shell are evidence for a past supernova event, where was the center of the explosion?  The presence of spectroscopic binaries may provide a clue.  These stars have  invisible companions that in several cases have been proven to be the skeletal remnant of an exploded star, either a neutron star or a black hole; see, for example, Escorza et al (2023) and Schmelz \& Verschuur (2022).

Fig. 4 shows a detailed view of the HI cavity at + 12 km s$^{-1}$ and  its surrounding shell of HI with the location of three spectroscopic binaries (SBs) found in the SIMBAD data base, indicated by red diamonds in Fig. 4.  These are amongst the prominent members of the USco Association.  At the left is $\nu$ Sco, a single-line SB, at a distance of 142 pc.  This may be the most likely candidate for the source of  the supernova event.  The center SB in Fig. 4  is $\beta$ Sco at a distance of 123 pc, but it is a 5-component, multi-star system without a clear indication in the literature that the SB category is substantiated.  The third SB, at the right in Fig 4, is $\delta$ Sco, about which much has been published.  For example, Miroshnichenko et al. (2013) derived an accurate distance of 136 pc but make no mention of its possible SB nature.  They did conclude that it might be part of a triple system with the third member yet to be detected.

\begin{figure}
\figurenum{4}
\epsscale{0.9}
\plotone{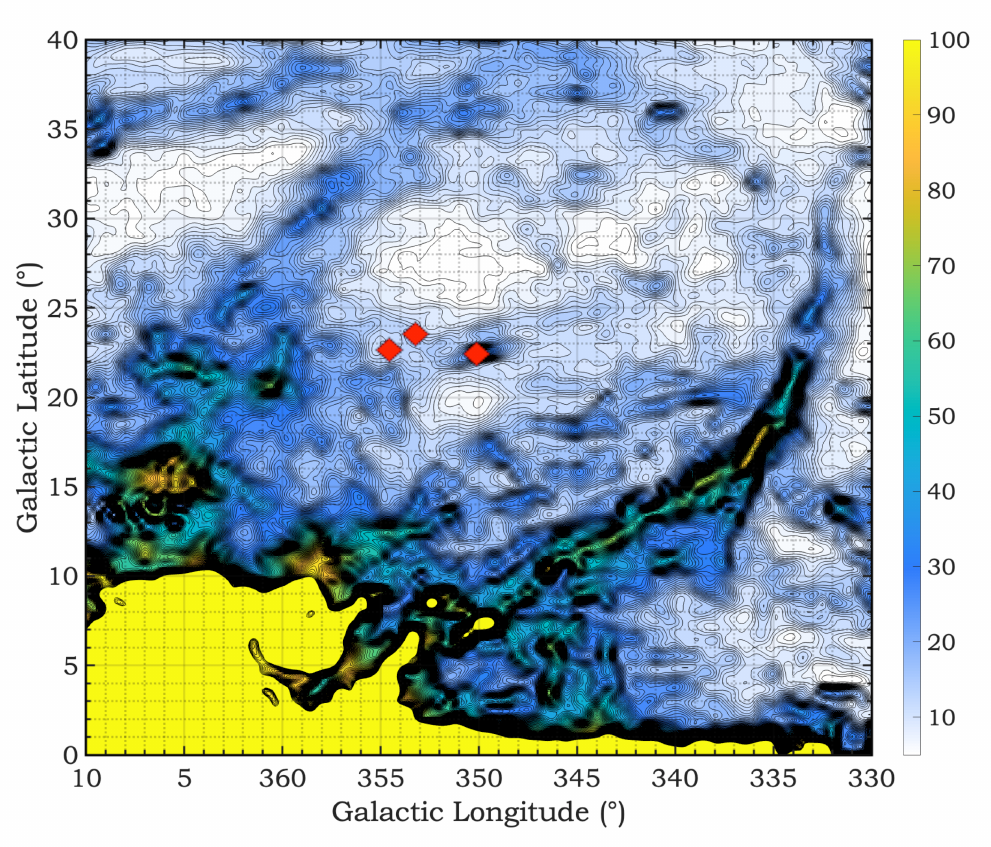}
\caption{\textit{LAB} \textit{l-b} data in a 2 km s$^{-1}$ wide band showing the HI structure at +12 km s$^{-1}$ toward the USco Association.  The locations of three spectroscopic binaries discussed in the text are shown as red filled diamonds. Legend in K. km s$^{-1}$.}
\end{figure}

But are the WD and its two mystery HI features located inside the cavity in depth?  To answer this question requires an estimate of the linear extent of the cavity.

A narrow range of distance estimates for the USco Association can be found in the literature.  Guo et al. (2025) give the distance to what they term the US region of the Scorpius OB Association of 147 pc.  Damiani et al. (2019) call the same region USC and report two distinct compact stellar populations with parallaxes of 7 and 6.5 to 6, which implies distances of 143 and from 163 to 166 pc.   de Geus et al. (1989) estimated the distance to be 158 pc.  

The width of the USco shell is about 30\arcdeg\ in latitude (Fig. 4).  If it is as deep as it is wide and using the Guo et al. (2025) GAIA-based distance estimate of 147 pc, its linear width is of order 84 pc.  This implies that  the cavity may extend from 105 to 187 pc in depth along the line-of-sight.  This places the WD, at its distance of 110 pc, inside or just at the edge of the USco cavity.  

There is more information related to the  depth of the USco cavity, as can be seen in Fig. 5, which shows a selection of four HI area maps based on \textit{LAB} \textit{l-b} map in 5 km s$^{-1}$ bands at the velocities indicated.  At +8 km s$^{-1}$ the shell seen so clearly in Fig. 4 is still prominent.  Its signature can also be recognized at +28 km s$^{-1}$.  There is even a hint of its  presence at +32 km s$^{-1}$ (not shown here).  At -4 km s$^{-1}$ the near side of the shell begins to fill in while the outline of a loop that is smaller than the one at +8 km s$^{-1}$ is intriguing.  It hints at a distinctly different driving force to create that structure.  At -12 km s$^{-1}$ a band of HI dominates the image.  This may be HI on the near-side face of the shell outlning the USco cavity. 

\begin{figure}
\figurenum{5}
\epsscale{1.0}
\plotone{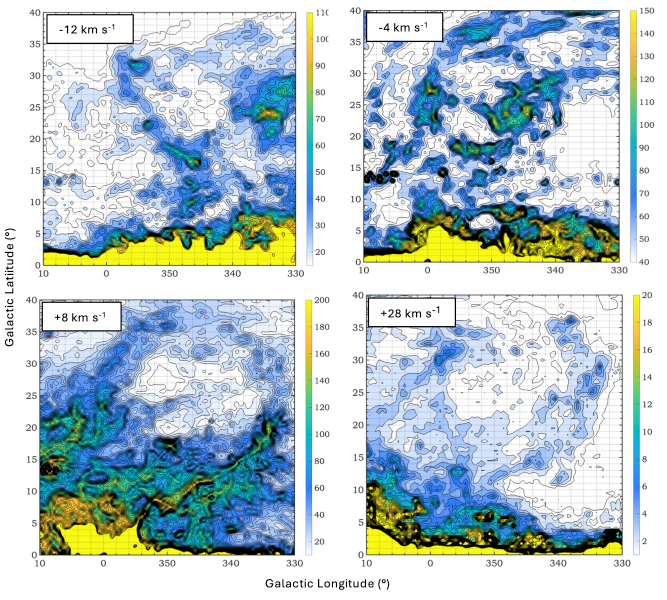}
\caption{\textit{LAB} \textit{l-b} data in 5 km s$^{-1}$ wide bands at the velocities indicated showing the outlines of the shell of HI surrounding the USco Association.  At +8 km s$^{-1}$ the shell is as clearly visible as it is in Fig. 4.  Its outlines are also  distinguishable at +28 km s$^{-1}$.   At -4 km s$^{-1}$ its outlines can still be seen and in addition there is evidence for a smaller ring of gas inside the larger one. The excess structure in the center of the area may include gas at the near-side face of the shell surrounding the USco Association.  At -12 km s$^{-1}$ a hint of structure related to the main shell is visible. Legends are in K. km s$^{-1}$. }
\end{figure}

Together, the morphologies seen in Fig. 5 at positive velocities suggest that the cavity is open to inter-arm space and is may be tunnel-like in shape.  The volume occupied by the USco cavity is therefore significantly larger than that implied by a simplistic interpretation of its appearance in an image at one velocity, such as Fig. 4.  We proceed under the assumption that the SB and its twin HI features are located inside the USco HI cavity, a volume of space that has been swept clear of most of the HI by a past supernova.

 \subsection{Other HI associated with the SB}
Fig. 6 shows the HI brightness in \textit{l-b} maps based on \textit{HI4PI} data at four select velocities each covering a single channel of data (1.28 km s$^{-1}$ wide).  Superimposed on these figures are dashed red lines drawn to connect the bulk of the emission at the given velocity to the location of the WD, marked by a red star.  The HI at these velocities in the area around the WD is oriented along filament-like structures and the star appears to be the focus of the HI morphology immediately around it.  

\begin{figure}
\figurenum{6}
\epsscale{0.8}
\plotone{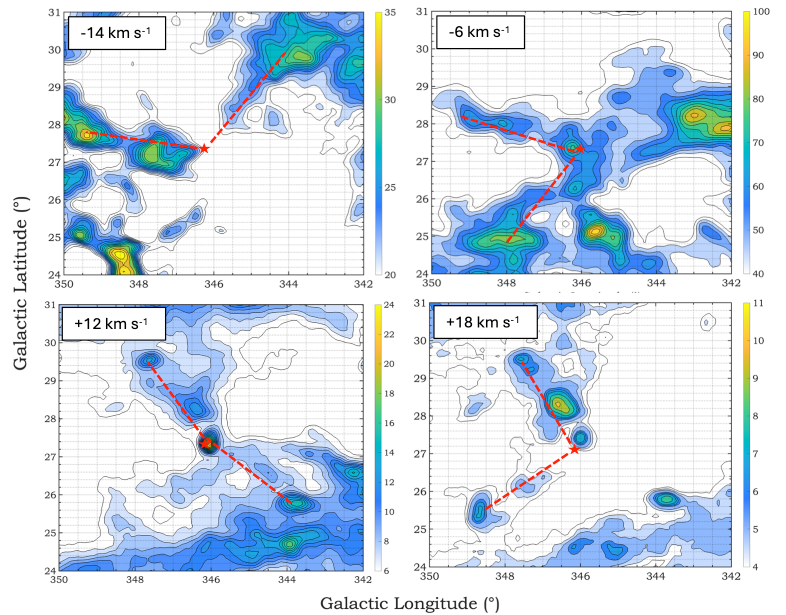}
\caption{\textit{HI4PI} \textit{l-b} data in a single channel, 1.28 km s$^{-1}$ wide, at the velocities indicated showing detailed views of the HI surrounding the white dwarf (red star symbol). Red lines show how the HI structure appears to be related to the location of the star.  This is strong circumstantial evidence that the star is the source of HI streams ejected in the past, probably during the planetary nebula phase in the star's evolution.  Legends are in K. km s$^{-1}$.}
\end{figure}

Examination of Fig. 6 suggests that the angular widths of the filament-like HI features are of order 0.\arcdeg5, which is about 1 pc at  the distance of the WD.  

To estimate the mass involved in the HI clouds shown in Fig. 2, two approaches were attempted.  In the first a composite spectrum was produced centered on the primary peak (Fig. 2a) for an area 2.4 pc$^{2}$ in extent (at the distance of the WD).  This spectrum was fitted with a Gaussian to estimate the amount of hydrogen at the far-side of the WD in this volume of space. The composite spectrum gives the average column density over the chosen area which allows the number of particles in that volume to be determined.  This led to a mass estimate of 0.07 M$_{\circ}$ around the peak of HI at far side of the WD.   

A second method considers the HI peak at -6 km s$^{-1}$ around (l, b) = (346.\arcdeg2, 27.\arcdeg3) and uses information on its \textit{contrast}  with respect to the background in the \textit{l-b} map, e.g. Fig. 2b, about 15 K over a velocity extent of order 4 km s$^{-1}$.  This  odd approach was necessitated because no unambiguous Gaussian decomposition could be obtained from the  HI profiles.  This leads to a mass for that feature of 0.06 Mo, comparable to the estimate for the peak at +12 km s$^{-1}$.

\section{Discussion}
The two unresolved (mystery) HI features shown in Fig.2  appear to be associated with a white dwarf star.   In the model considered here the feature at -6 kms$^{-1}$ is located on the near side and the +12 km s$^{-1}$ feature is at the far side of the star.  The additional presence of the filament-like structures of HI emission surrounding the WD seen in Fig. 6 suggests that we may be dealing with a very old planetary nebula (PNe) that has expanded into the cavity in interstellar HI created by a supernova (and/or stellar winds) in the USco OB Association.  The PNe gas has now cooled to the point that its hydrogen gas is neutral and radiating the $\lambda$21-cm spectral line.  

In what follows it is assumed that we are dealing with a very old, or relic, PNe.  The two unresolved HI peaks shown in Fig 2, the features that began this study, may represent gas ejected by the star on an axis aligned nearly along the line-of-sight.  In the absence of further HI data, the pair could be interpreted as an example of a bi-polar planetary nebula.  Such nebulae are well-known, e.g.  Clyne et al. (2015) and Ondratschek et al. (2022).  However, those examples concern white dwarf binaries, and we have no evidence regarding the possible binary nature of WD \#1 in Table 1.  

If this simple model of a relic PNe is valid there should be evidence for HI structure at the boundary of the ensemble of HI features described above.  HI emission from such a shell around the WD should be detectable at a velocity of approximately +3 km s$^{-1}$, the average between the two mystery features at + 12 and -6 km s$^{-1}$.  (The stellar velocity is not known at this time.)  However, not all PNe show simple structure such as is epitomized by the Ring Nebula.  For example, the Dumbbell Nebula (M27) exhibits two arcs on opposite sides of a white dwarf star that define the outer bounds of that PNe.  Also, examination of the Hubble Atlas of Nebulae or the Google collection of planetary nebula images make it clear that structure in PNe is far from iconic and that, instead, tangled messes of filaments and bright knots of emission inside the nebulae are the rule, not the exception.  NGC 5189 is a good example of a chaotic looking PNe.

Fig. 7 shows the HI structure based on the HI4PI in the general area of the WD at three velocities that reveal indications of boundary segments surrounding the star.  At +3 km s$^{-1}$ a distinct arc in the HI emission is seen to the north of the star.  If there is a similar feature to the south it, unfortunately, is lost in the unrelated foreground/background structure at the lower left of the map.  At -10 km s$^{-1}$ the ring sketched in Fig. 7 encompasses an arc of HI structure to the east (left) of the star.  Even at +10 km s$^{-1}$ HI emission peaks appear to define a slightly elliptical outline of a shell about the star.  If the HI structure immediately around the WD indicates the relics of an ancient PNe, it is clear that the nebula is not symmetric in outline on the sky.  (HI emission structure inside the red dashed outlines in Fig. 7 include the filamentary features shown in Fig. 6, \S2.5.)

\begin{figure}
\figurenum{7}
\epsscale{0.5}
\plotone{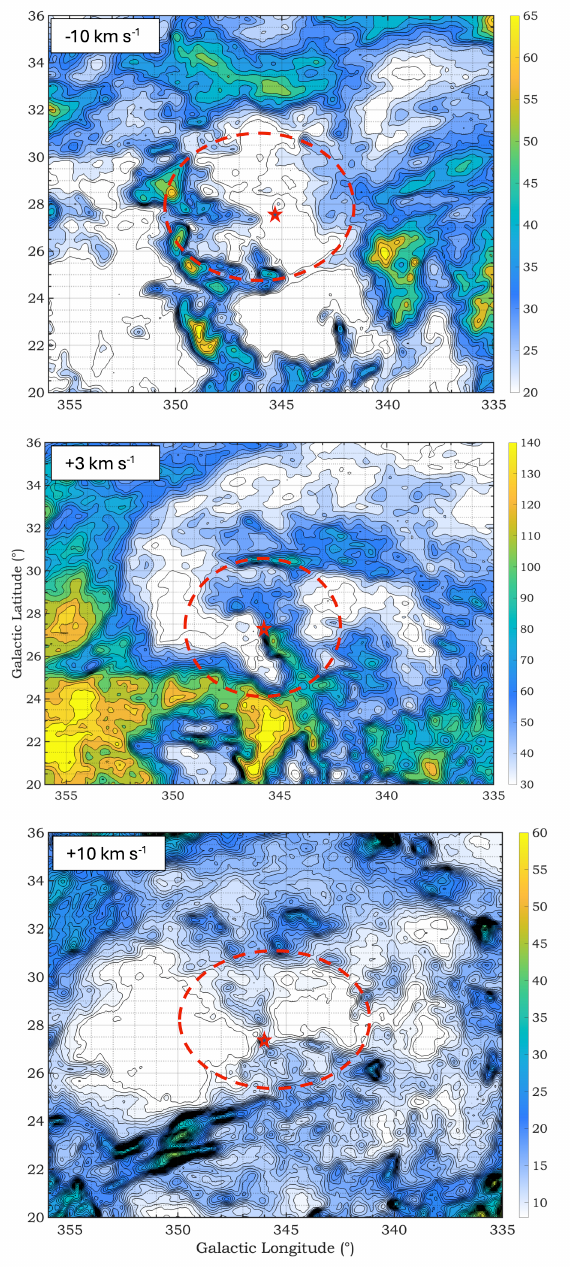}
\caption{\textit{HI4PI} \textit{l-b} data in a single channel, 1.28 km s$^{-1}$ wide, at the velocities indicated showing segments of  HI in arcs around the white dwarf (red star symbol) discussed in the text.  These present evidence for segments of possible rings, red dashed circles.  The ring signature can be recognized at  +3 and +10 km s$^{-1}$ while at -10 km s$^{-1}$ the HI structure fills in the gap seen at +10 km s$^{-1}$. Legends are in K. km s$^{-1}$. }
\end{figure}

Examination of Figs. 6 \& 7 and the other independent \textit{l-b} maps covering the full range of velocities in the area suggest the two mystery HI clouds cover about 10\% of the total area of the filaments.  This implies a total mass of order 0.7 M$_{\circ}$.   At a velocity of expansion of 20 km s$^{-1}$, based on an examination of  these figures, the energy of expansion of 6 x 10$^{45}$ ergs is derived.  The angular width of the ring defined by segments of HI seen in Fig. 7 is approximately  6\arcdeg, which corresponds to a radius of 5.8 pc.  At the expansion velocity of 20 km s$^{-1}$ the derived age is then about 3 x 10$^{5}$ years for the relic PNe.
 
Finally, why are the two features (\S2.20) located at the star's position unresolved while the filament-like structures in Fig. 6 are clearly broader than the beam width?  A possible explanation is illustrated in the \textit{l-b} map at +2 km s$^{-1}$, Fig. 8.  If the line-of-sight toward the star were along the axis of such a structure (red dashed line in  Fig. 8), the HI profile would exhibit a relatively narrow emission line superimposed on a broader underlying component, broader in angle and velocity, with the relative brightness and angular extent of the components determined by morphology along the line-of-sight.  

\begin{figure}
\figurenum{8}
\epsscale{0.5}
\plotone{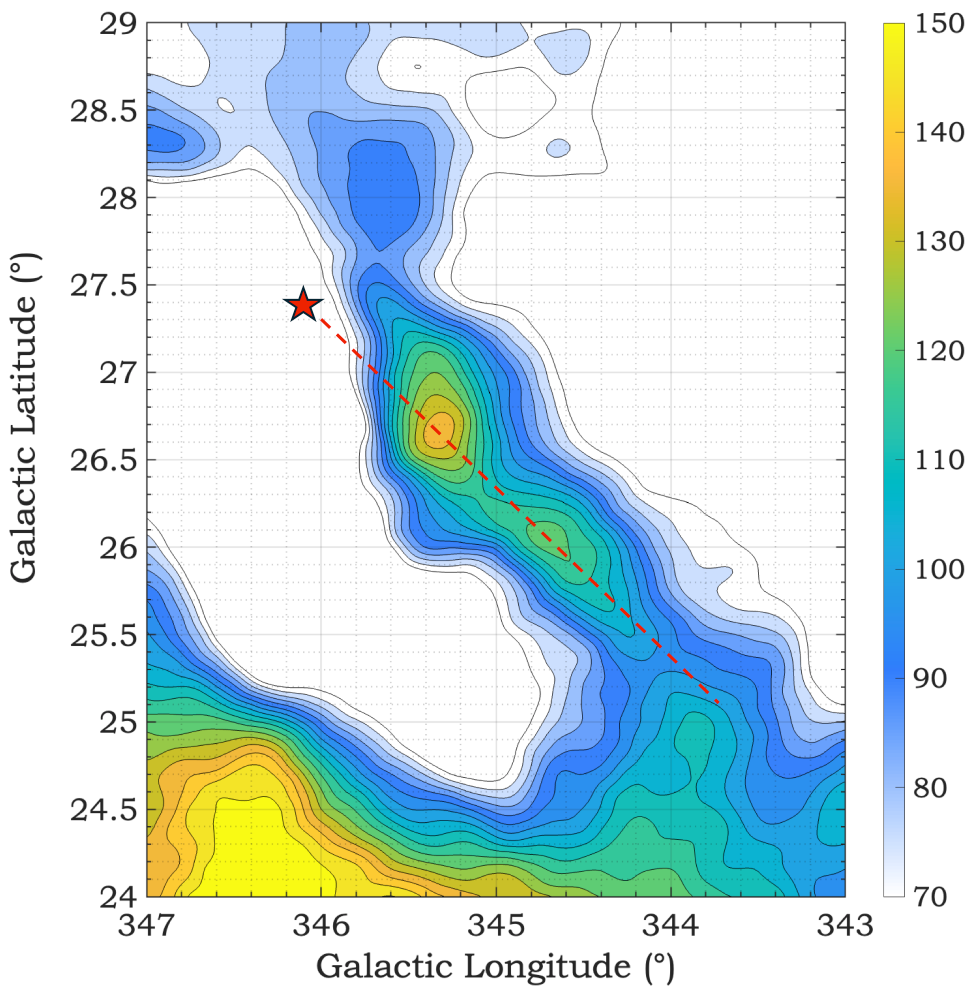}
\caption{\textit{HI4PI} \textit{l-b} data, one channel wide, showing a detailed view of a linear filamentary HI structure radiating away from the white dwarf (red star) at a velocity of +2 km s$^{-1}$.  Seen along the line-of-sight simulated by the overlaid red dashed line an association between the star and a small angular diameter peak might be inferred, see text. Legend is in K. km s$^{-1}$.}
\end{figure}

This report has been qualitative in nature, but the above considerations present enough circumstantial evidence to reinforce the suggestion that the HI structures in the immediate vicinity of the WD (as projected on the sky) are the relics of an ancient PNe.  This is confirmed by the limited quantitative data that have been alluded to above.  For example, the rough mass estimate for the HI  immediately around the WD of 0.7 M$_{\circ}$ (above) is of the order of magnitude associated with planetary nebulae, 0.1 to 1 M$_{\circ}$.  Also, typical planetary nebulae have expansion velocities of from 20 to 50 km/s and the implied energies of expansion are between 4 x 10$^{44}$ and 2 x 10$^{46}$ ergs. The value of  6 x 10$^{45}$ ergs for the hypothesized relic PNe derived above lies within this range.  These derived values are consistent with the relic PNe hypothesis.  

No clear one-to-one relationship between IRAS 100$\mu$ dust morphology and that of the filamentary HI features seen in Fig. 6 was found.  On the other hand, the arc of HI seen at latitude 30\arcdeg\ in Fig. 7 at +3 km s$^{-1}$ appears to show 100$\mu$ emission.  The detailed relationship between the complex HI structure discussed here and the even more complex 100$\mu$ dust features in the area probably deserves further study.  Observations of CO in the region by de Geus et el. (1990) did not extend beyond the Ophiuchus region encompassed by the red rectangle in Fig 1, which means that no CO data are available for the putative relic PNe discussed above.

\section{Conclusions}
The goal of this paper was to draw attention to the presence of  an unusual pair of apparently unresolved HI clouds that appear to be associated with a white dwarf star at the same position.  The ensemble is located inside a cavity surrounding an association of OB stars, the Upper Sco-Cen OB2 association, which has been well-studied by others for several decades.

The  other HI structure immediately around the white dwarf star carry the hallmarks of what may be a 3 x 10$^{5}$ year old  planetary nebula that has cooled sufficiently so that the outlines of its original structure can now be recognized in area maps of $\lambda$21-cm emission from neutral hydrogen.  In general an  old planetary nebula would cool and become mixed with surrounding interstellar HI, but in this case the nebula expanded into a relative void that is the cavity surrounding the Upper Sco-Cen OB2 association.

Such a situation would only be detectable in directions where there is little ambient HI present, for example at larger distances from the Galactic disk.  In the case reported here, a unique set of circumstances allows the detection of this phenomenon at a relatively low Galactic latitude because it is found inside the cavity surrounding the nearby Sco-Cen OB2 Association.

\section{Acknowledgments}
I owe a debt of gratitude to Dr. W Butler Burton for encouragement and useful discussions, and for drawing my attention to mportant references.  I thank  Dr. Ana Escorza for pointing me toward literature that confirmed the nature of the white dwarf star discussed in this study, Dr. A. Miroshnichenko for useful information, and Dr. Tom Dame who willingly helped when unexpected software glitches prevented me from making further progress in data display issues in work such as this.  An anonymous reviewer is especially thanked for pointing out key aspects of planetary nebulae that had to be taken into account in order to understand the data.


\begin{thebibliography}{}
\bibitem[Ben Bekhti et al. (2016)]{Bekhti16}Ben Bekhti, N., Fl\'oer L., Keller, R., et al. 2016, A\&A, 594, A116
\bibitem[Clyne et al. (2015)]{Clyne15}Clyne, N., Akras, S., Steffen, W., et al. 2015, A\&A, 582, A60 
\bibitem[Damiani et al. (2019)]{Damiani19}Damiani,  F., Prsinzano, I, Pillitteri, G, et al. 2019, A\&A, 623, A112
\bibitem[de Geus et al. (1989)]{de Geus89}de Geus, E. J., de Zeeuw, P. T., \& Lub, J. 1989, A\&A, 216,44
\bibitem[de Geus et al. (1991)]{de Geus91}de Geus, E. J., \& Burton, W. B. 1991, A\&A, 246, 559
\bibitem[de Geus et al. (1992)]{de Geus92}de Geus, E. J. 1992, A\&A, 262, 258
\bibitem[de Geus et al. (1990)]{de Geus90}de Geus, E. J., Bronfman, I., \& Thaddeus, P. 1990, A\&A, 231, 137
\bibitem[Escorza et al. (2023)]{Escorza23}Escorza, A., Larinkuchi, D., Jorissen, A., et al. 2023, A\&A, 670, L14
\bibitem[Gentile-Fusillo et al. (2021)]{Gentile-Fusillo21}Gentile-Fusillo, N. P., Tremblay, P.-E, Cukanovaite, E., et al.  2021, MNRAS. 508, 3877
\bibitem[Guo et al. (2025)]{Guo25}Guo D., Kaper, L., Brown, A. G. A., et al. 2025, A\&A, 2025, 696, A119
\bibitem[Kalberla et al. (2005)]{Kalberla05}Kalberla, P. M. W., Burton, W. B., Hartmann, D., et al. 2005, A\&A, 440, 775
\bibitem[Kerp et al. (2011)]{Kerp11}Kerp, J., Winkel, B., Ben Bekhti, N., et al. 2011, Asron. Nachr, 332, 637
\bibitem[Krause et al. (2018)]{Krause18}Krause, M. G. H., Burkert, A., Diehl, R., et al.  2018, A\&A, 619, A120
\bibitem[McClure-Griffiths et al. (2009)]{McClure-Griffiths09}McClure-Griffiths, N. M., Pisano, D. J., Calabretta, M. R., et al. 2009. \apjs, 181, 398
\bibitem[Miroshnichenko et al. (2013)]{Miroshnichenko13}Miroshnichenko, A. S., Pasechnik, A. V., Manset, N., et al. 2013, \apj, 766,119 
\bibitem[Olano et al. (1981)]{Olano81}Olano, C. A., \& P{\"o}ppel. W. G. L. 1981, A\&A, 95,316
\bibitem[Ondratschek et al. (2022)]{Ondratschek22}Ondratschek, P. A., R\"opke. F. K., Schneider, F. R N., et al. 2022, A\&A, 660, L8 
\bibitem[Poppel et al. (2022)]{Poppel10}P{\"o}ppel, W. G. L., Bajaja, E., Arnal, E. M., \& Morras, R. A.  2010, A\&A, 512, A83
\bibitem[Schmelz et al. (2022)]{Schmelz22}Schmelz, J. T., \& Verschuur, G. L. 2022, \apj, 938, 68
\bibitem[Winkel et al. (2016)]{Winkel16}Winkel, B., Kerp, J., Fl\'oer, L., et al. 2016, A\&A, 585, A41 
\end{thebibliography}
\end{document}